# EEG Emotion Recognition From AI-Generated Biodigital Architecture Images

Hongye Yang[1], Eva Guttmann-Flury[2]

[1]Beijing Institute of Architectural Design Co., Ltd., Beijing, China, [2]Shanghai Jiao Tong University,Shanghai, China

yanghongye@biad.com.cn, hyang783@gatech.edu ; eva.guttmann.flury@gmail.com

**Abstract.** Emotional responses to biodigital architecture were examined using electroencephalographic (EEG) data from AI-generated images. A pre-experiment involving 336 participants identified 60 images, selected from an initial pool of 600, that elicited strong emotional responses categorized as awe, disgust, or content. These images were used for EEG recordings of 52 volunteers, with channel selection and sample size estimation based on the analysis of an existing dataset. Gamma and delta bands yielded the highest classification accuracy, with the gamma band achieving 77.07% ± 13.8% accuracy for the awe emotion. Key factors such as greenery and non-uniform granularity were linked to positive emotions, while dampness triggered negative reactions. These results emphasize the significance of incorporating natural elements and varied textures in biodigital architecture to enhance aesthetic appeal and acceptance. The study demonstrates EEG's capability to objectively assess architectural preferences, providing valuable insights for architects to design engaging and sustainable environments.

## 1 Introduction

Biodigital architecture fuses biological and digital technologies, using bio-genetic materials for self-organizing buildings (Estévez, 2009). This form aims for sustainability through optimized design and material use (Abdallah and Estévez, 2021). Since the first biodigital architecture lab at the International University of Catalonia in 2000, various projects have demonstrated this emerging style's potential (Estévez, 2009).

Biodigital architecture offers energy-efficient, low-carbon solutions, as seen in projects like the "Pod Hotel" (Dollens, 2009) and the "Genetic Barcelona Project"(Estévez and Navarro, 2017). It also shows promise for social housing, with prefabricated units and cost-effective "Biodigital Barcelona Clay Bricks"(Abdallah and Estévez, 2021). Additionally, it introduces innovative design methodologies, such as the "architectural bark" irrigation system (Cruz and Beckett, 2016) and microbial and moss facade systems (Mao, 2022).

Research on the aesthetics and acceptance of biodigital architecture is still limited. With few global examples, it is difficult to assess public appreciation or resistance. This situation mirrors early modernist architecture, which was criticized in the 1980s for being cold and impersonal (Anon., 1988). Urban designers need to integrate biodigital architecture into contemporary cities in a way that is acceptable to the public.

User preferences, typically gathered via self-reported questionnaires, can distort conclusions about emotional responses to designs (Hulliyah, Awang Abu Bakar and Ismail, 2017). To address this, cognitive states can be directly measured using electroencephalographic (EEG) signals, offering greater objectivity and accuracy than self-reported metrics (Hulliyah, Awang Abu Bakar and Ismail, 2017).

Research shows distinct neural responses to emotions in specific cortical regions (Yang, Liu and Xie, 2026): the medial prefrontal cortex aids in emotional decision-making(Yao et al., 2024), the orbitofrontal cortex evaluates emotionally relevant scenarios, and the posterior cingulate cortex processes emotionally significant stimuli and memory (Phan et al., 2002). Additionally, visually salient features influence (e.g., color, shape) neural activity in visual areas (Etkin, Büchel and Gross, 2015).

Neuroscience can also enhance living environments to meet human needs (Adityo, 2024). EEG has been used to assess indoor design preferences (Kim, Cheon and Kang, 2019) and guide outdoor nighttime lighting (Chang, Dong and Jun, 2020). Its real-time capabilities enable innovations in architectural design, such as adjusting color and window size based on feedback (Zhang, Li and Guo, 2023). However, EEG's use in biodigital architecture remains largely unexplored. Our goal is to provide methodological insights for future research to align biodigital architecture with public aesthetic preferences.

This study aims to provide practical insights for architects and designers in selecting an affordable EEG headset. We hypothesize that, despite potential lower accuracy, a rigorous methodology will yield robust conclusions. Due to the limited real-world examples of biodigital architecture, we will use AI-generated images to evoke strong emotional responses and valuable insights.

## 2 Methodology

To ensure the reproducibility of this study, we highlight critical experimental design decisions. Electrode selection in EEG experiments is pivotal, and determining an appropriate sample size is crucial for accurately capturing the intended phenomenon. We identified the "imagined emotion study" dataset (Onton and Makeig, 2022) as particularly relevant, guiding our selection of EEG channels and the minimum number of subjects needed. Our investigation focuses on identifying three primary emotional states from this high-density EEG dataset: awe (positive), disgust (negative), and content (neutral).

### 2.1 Image selection

Geometric forms in architectural images (e.g., scale, proportion, curvature) can affect emotional states as measured by EEG (Shemesh et al., 2021). Additionally, biodigital architecture, being less familiar to the public, serves as an effective stimulus for eliciting emotional responses measured by EEG (Thammasan et al., 2017). Due to the scarcity of real-world examples, we generated 600 AI-created images themed around biodigital architecture. Initial examples were sourced from social media, descriptions were crafted using ChatGPT-4 (OpenAI et al., 2023), and DALL-E generated images based on emotional states like awe, disgust, and contentment (Betker et al., n.d.).

Participants rated the images on an online platform using emojis, from a crossed-out eyes face (strongest negative reaction, scored as 0) to a star-struck face (strongest positive reaction, scored as 4). The questionnaire can be found at https://git.com//QinX/bio-experiment. Images were categorized based on emotional responses: awe images had high positive scores, disgust images had high negative scores, and content images had minimal score variability. This approach mitigated central tendency bias and ensured a broader distribution of responses for new participants.

### 2.2 Channel selection

Channel selection was guided by the analysis of the "imagined emotion study" dataset(Onton and Makeig, 2022). Source localization was conducted using the eLORETA (exact Low Resolution Electromagnetic Tomography) method with default parameters in the MNE library in Python(Gramfort, 2013). For each subject, the noise covariance matrix was estimated from pre-stimulus data to ensure accurate identification of neural activity. The fsaverage template brain was used to calculate the forward model, providing a standardized framework for analysis（see figure 1）.

Source localization plots were then generated and compared across subjects. Brain areas consistently activated across multiple subjects were identified as regions of interest for our experiment. Figure 1 illustrates the inter-subject variability of the activated brain regions. Consistent with previous research, this analysis revealed that the occipital, parietal, and medial prefrontal cortices are implicated in emotional responses(Barbas, 2000, Etkin, Egner and Kalisch, 2011). Based on these findings, we selected the EEG recording sites Fp1, Cz, CPz, CP1, CP2, Pz, O1, and O2 to ensure comprehensive data collection from these regions.

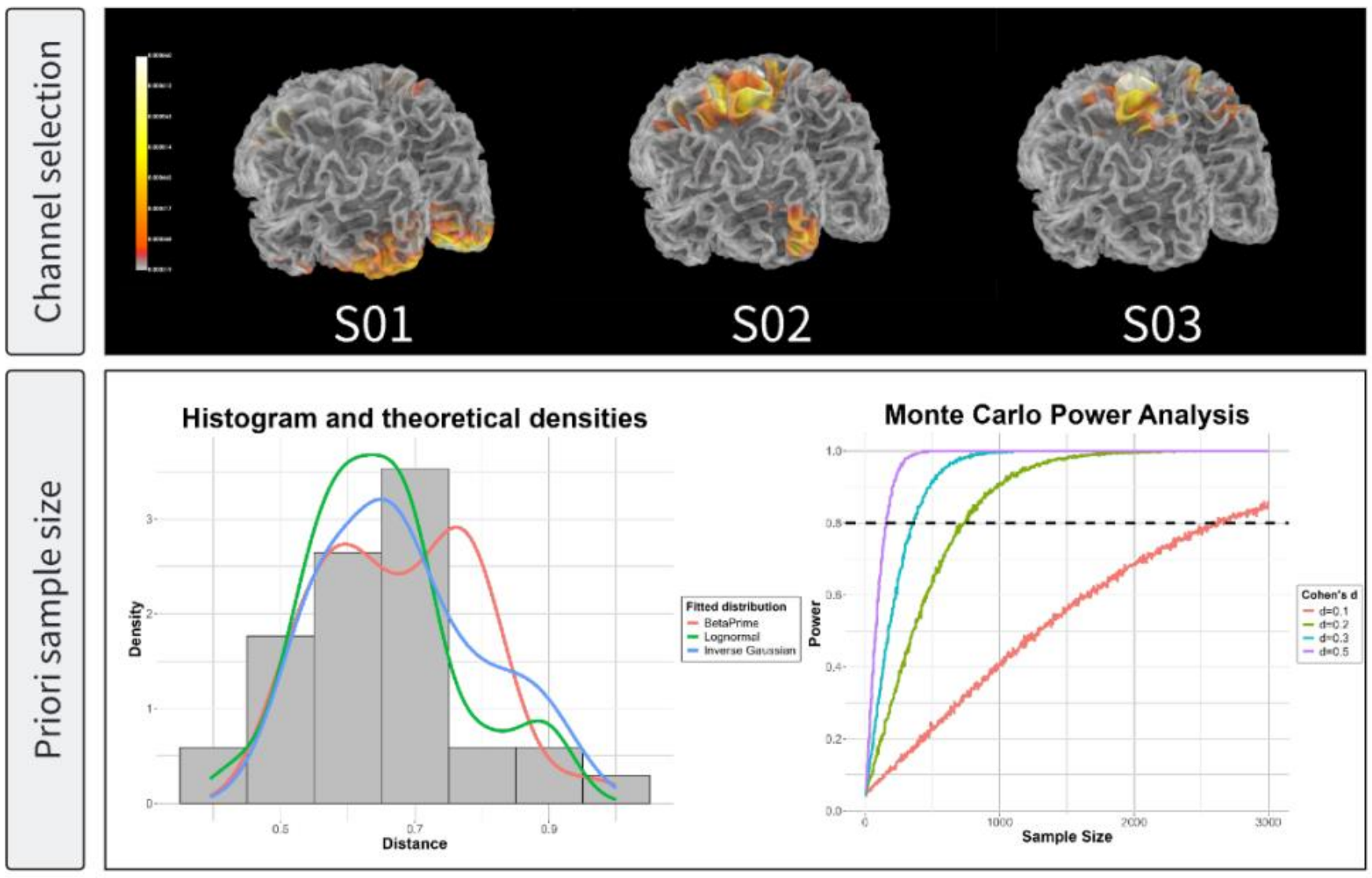

**Figure 1.** Estimating channel locations and a priori sample size from the online-available dataset "Imagined emotion study". Source: Authors, 2024.

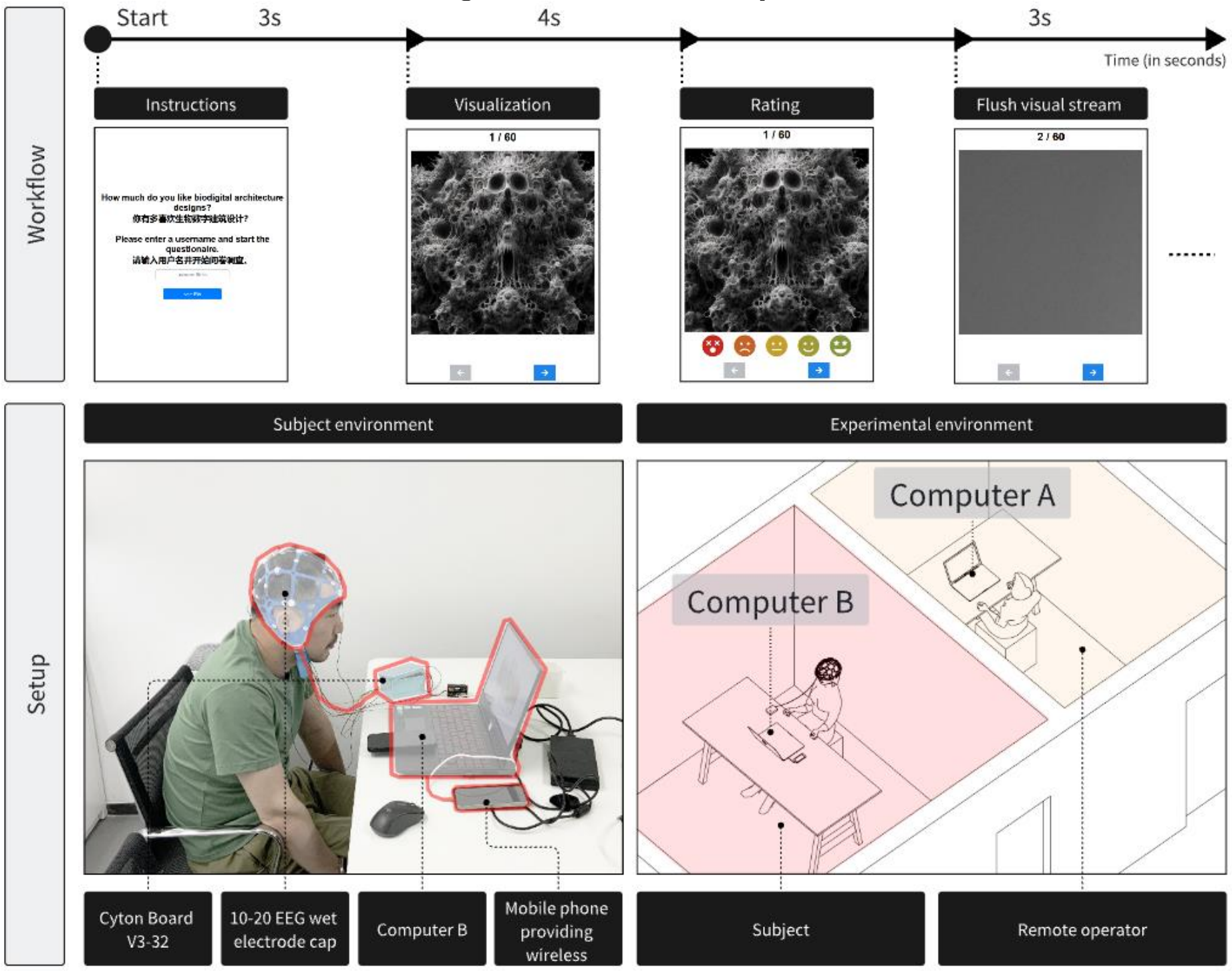

**Figure 2.** Experimental workflow and design. Source: Authors, 2024.

### 2.3 A priori sample size estimation

Sample size was estimated using covariance matrices from the "imagined emotion study" data at eight electrodes during a 1-second interval following emotion-inducing stimuli. Riemannian averages, medians, and distances were calculated, allowing us to estimate Cohen's d by comparing emotion-related and non-emotion-related stimuli. The resulting Cohen's d values were approximately 0.1 for content, 0.2 for disgust, and 0.15 for awe, indicating stronger brain reactions to extreme emotions, particularly negative ones.

Each distance distribution was fitted to 30 models, and the top three were visually compared with the empirical data, showing the Inverse Gaussian distribution was the best fit for the content emotion. Through the Fitted-Distribution Monte Carlo (FDMC) approach, samples were generated from this fitted distribution to estimate the necessary sample size for the EEG experiment (Guttmann-Flury et al., 2019a)（see figure 1）. To keep sessions under 30 minutes, each was limited to 20 trials per emotion. Conservative estimates required 2,585 trials (43 subjects), while balanced estimates suggested 730 trials (12 subjects).

### 2.4 Participants

This study follows the Declaration of Helsinki principles and included 52 healthy volunteers (24 women, 28 men; mean age 30.62 ± 7.52), all of whom gave written informed consent for anonymized data use. None had prior exposure to AI-generated biodigital architecture images, ensuring a novel experience. Prior to the experiment, participants filled out a questionnaire covering demographics, physiological data, and psychophysiological states. Handedness was not a disqualifying factor. They were briefed on the experiment and asked to stay still and quiet during the image presentations.

### 2.5 Paradigms

The experiment involves collecting two types of data simultaneously: participant responses to a biodigital architecture image questionnaire and concurrent EEG signals, both synchronized using shared time markers and stored on computer B (see Figure 2).

Each trial begins with a 4-second display of a randomly chosen image, during which participants remain still and silent. The image onset is marked as "IMAGE" with its corresponding number in the Image column. Afterward, participants choose an emoji representing their emotional response, from "strongly dislike" to "strongly like," recorded as "EMOJI" in the Trigger column.

Following this, a gray image is displayed for 3 seconds to reset the visual stream before the next image. The questionnaire consists of 60 images, with each session lasting about 10 to 15 minutes. Participants may speak but must avoid physical movement during selection to prevent EEG artifacts.

### 2.6 Acquisition

The Cyton Board V3-32 by OpenBCI, equipped with a WiFi Shield, was used for the EEG experiment, enabling data transmission and storage via WiFi. Participants wore a 10-20 wet electrode cap and were seated in an isolated room. Both EEG data and questionnaire responses were stored on a designated computer (Computer B, see Figure 2).

Due to environmental constraints, a wired network connection was not feasible, so mobile phone USB tethering provided network access. Although this may introduce some electromagnetic interference, it is expected that this noise will be more manageable during preprocessing compared to artifacts

### 2.7 Preprocessing

Raw EEG data were down-sampled to 250 Hz. Spike artifacts from pulsed electromagnetic signals were identified by comparing the down-sampled data against a fourth-order 5 Hz low-pass Butterworth filter. Extreme values above a 200 µV threshold were replaced with their filtered values. The data were then low-pass filtered with a $4^{th}$-order 15 Hz Butterworth filter, using either the original down-sampled or de-blipped data, depending on the detected spikes.

Blinks were removed using the Adaptive Blink Correction and De-Drifting (ABCD) algorithm (Guttmann-Flury et al., 2019b). A moving average window detected threshold-exceeding data at the fronto-polar channel FP1, and potential blinks were assessed by evaluating pre- and post-amplitude. FP1 signals were compared with the central electrode CZ to exclude brain activity or electrode dislodgment. Identified blinks were categorized by peak amplitude, and class-dependent averages were computed. Blink-Related Potentials (BRPs) were assessed automatically to discard data lacking typical blink patterns. Five subjects were excluded: S04, S05, S25, and S28 for faulty FP1 electrodes, and S21 for faulty O1 and O2 electrodes.

Drift curves for each channel were computed using a $4^{th}$-order 2 Hz low-pass Butterworth filter to reveal the global trend and were removed from the raw data, along with the appropriate class-dependent smoothed Grand Median BRP fitted to each blink. This procedure ensured the acquisition of artifact-free EEG data while preserving the original spectral characteristics.

### 2.8 Processing

#### 2.8.1 EEG

EEGNet is a specialized convolutional neural network tailored for EEG-based Brain-Computer Interfaces (Lawhern et al., 2018). It employs depthwise and separable convolutions to efficiently capture EEG features while maintaining a compact architecture. Depthwise convolutions filter individual channels independently, while separable convolutions enhance feature extraction by decomposing operations into depthwise and pointwise convolutions. This design enables effective processing of EEG data even with limited training and computational resources, making it suitable for real-time BCI applications.

The EEGNet model in this study has three blocks. The first block features a temporal convolution layer for frequency-specific pattern capture and a spatial batch normalization for training stability. The second block uses depthwise convolution to apply spatial filters to individual channels and includes batch normalization to enhance learning. The third block incorporates separable convolution, combining depthwise and pointwise convolutions for better feature extraction and dimensionality reduction. The output is flattened and processed through a fully connected layer with a sigmoid activation function for classification. The model is trained with binary cross-entropy loss and the Adam optimizer, with backpropagation for predictions and weight updates.

#### 2.8.2 Image

Each image is first segmented into a 4x4 grid, followed by K-means clustering to identify the three primary colors in each segment. Pairwise Euclidean distances between the primary colors across all segments are computed to measure color similarity. The most distinct colors are chosen based on the distance matrix, yielding three principal colors for each image. Visual inspection confirms the most discriminative color, which is used as the reference for further processing.

After segmentation and color analysis, each image is evaluated using ChatGPT-4 for its characteristics in six aspects: dryness, greening, curvature, smoothness, brightness, granularity, and texture. Each characteristic is scored from 1 to 5, with 1 representing the lowest and 5 the highest degree.

## 3 Results

### 3.1 Image selection

During the pre-experiment phase, 336 participants were recruited, each allowed to interrupt the questionnaire at will. Only those who classified a minimum of 50 images were included in the final dataset, resulting in 259 participants, who classified an average of 149 images each. The most discriminative images were grouped into three emotions: awe, disgust, and content, as described in the methodology section. Figure 3 displays the emotion score distribution across these categories and highlights the top two representative images for each.

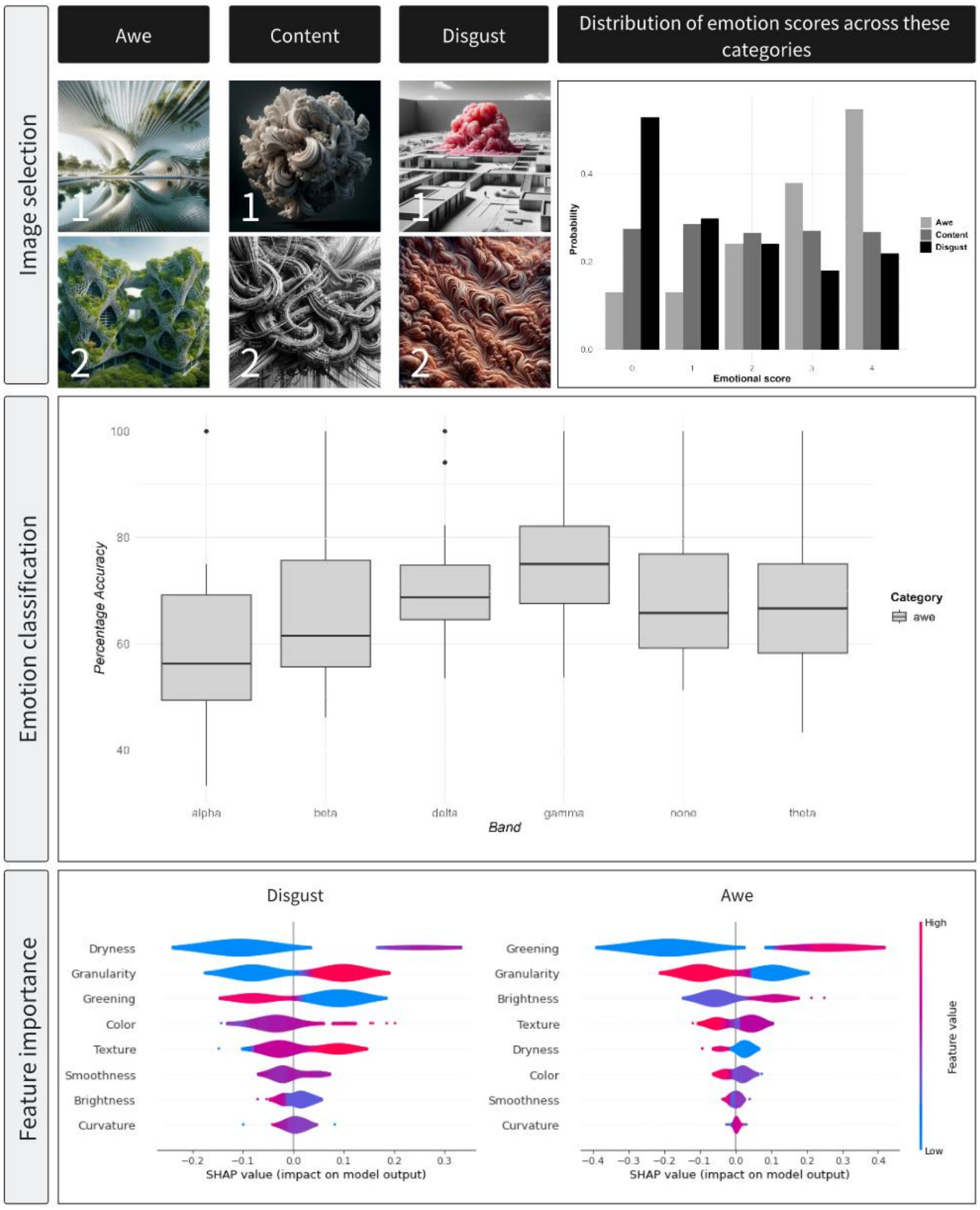

**Figure 3.** Results from the pre-experiment to identify 60 images, the classification accuray across filtering bands for the awe emotion, and the feature importance order for awe and disgust. Source: Authors, 2024.

## 3.2 Emotion classification

EEGNet was applied to the ABCD cleaned data across five frequency bands — delta (0.5-4 Hz), theta (4-7.5 Hz), alpha (7.5-12.5 Hz), beta (12.5-30 Hz), and gamma (30-45 Hz) — using a $4^{th}$-order Butterworth filter, as well as to unfiltered data. Results presented in Figure 3 focus on the "awe" emotion for two reasons. First, "disgust" elicited minimal neural responses, failing to exceed the 33% threshold across frequency bands, possibly due to the unfamiliar architectural shapes of the images. Second, "content" serves as a reference class that does not represent "awe" or "disgust," risking over-representation and bias. The analysis of "awe" revealed higher accuracy in the gamma and delta bands across subjects compared to the other frequency bands. The analysis indicated higher accuracy for "awe" in the gamma (77.07% ± 13.80%) and delta (70.65% ± 10.96%) bands compared to other frequency bands (see Figure 3).

### 3.3 Feature importance

To clarify the features affecting the model's output, SHAP values (SHapley Additive exPlanations)—a method from cooperative game theory—were used to improve machine learning model transparency. Categories from gamma band classification helped rank feature importance across emotions. As expected, the presence of trees and plants was a primary factor in preferences for certain architectural designs. Image granularity also played a significant role; disliked images were mainly attributed to perceived dryness, with "sticky" images receiving negative responses. Previous studies indicate that the sensitivity of the suprachiasmatic nucleus (SCN) to light affects preferences, favoring buildings with strong light-dark contrasts (Eberhard, 2009). The current research also confirms this tendency, suggesting that images with overly uniform light and dark contrasts are less popular.

## 4 Discussion

### 4.1 Summary of results

The gamma (77.07% ± 13.80%) and delta frequency bands showed the highest explanatory power in predicting participants' preferences for biodigital architecture images. Awe was linked to the presence of trees and plants, consistent with prior research on brain activity in the monkey visual cortex when exposed to natural scenes (Whittingstall and Logothetis, 2009).

Due to people's preference for nature-inspired living environments, biodigital architecture with natural styles can enhance mental health (Adityo, 2024). Although granularity and brightness had lower predictive power than greenery, they still influenced preferences, with non-uniform and well-lit images generally favored due to their visual complexity and illumination. Conversely, images with "stickiness" or a damp, unclean appearance were consistently disliked, likely due to evolutionary instincts associating such environments with health risks, aligning with environmental psychology and neuroscience theories.

## 4.2 Limitations

The Cyton Board V3-32, equipped with a WiFi Shield is susceptible to electromagnetic interference, which is exacerbated by using a mobile phone's USB interface for network connectivity. This introduces significant noise to the EEG signals. Additionally, the experimental setting lacks professional laboratory control, leading to variability from human activities. Despite efforts to limit movement, subtle participant actions, muscle contractions, or verbalizations before emoji selection can create artifacts that affect EEG data precision.

These artifacts may not be fully addressed by the algorithms in use, and limitations in computational power or algorithm complexity could further impact the data's accuracy. Image processing is also constrained by resolution and quality, which may result in the loss of intricate color and shape features. Moreover, the sample size analysis, based on a different dataset, may not account for differences in EEG electrode density or the experimental environment, potentially limiting the generalizability of the results for biodigital architecture design.

## 4.3 Conclusions

Integrating EEG data to understand emotional responses to biodigital architecture enables designers to align their creations more closely with user preferences. The use of EEGNet and SHAP values effectively identifies visual elements that evoke positive or negative emotions, such as greenery and visual complexity, offering architects actionable insights for enhancing aesthetic appeal. Notably, EEG power in the gamma (30-45Hz) and delta band (0.5-4 Hz) are significant predictors of the awe emotion, underscoring their relevance in visual processing of natural environments and color discrimination. The study reveals that perceived dryness or "stickiness" leads to negative reactions, emphasizing the need for materials and designs that convey cleanliness and health. Conversely, higher granularity — referring to more detailed and textured surfaces — elicits more favorable responses, suggesting a preference for visually complex and rich designs. Incorporating granular textures in biodigital architecture can enhance the sensory experience, making spaces more dynamic and engaging. Additionally, this style also contributes to sustainability, improving humanity's capacity to cope with climate change.

Future research can build upon these findings by broadening the image dataset to include a more diverse array of biodigital architectural styles and environmental contexts, thus improving feature analysis and the generalizability of results. Exploring the impact of various lighting conditions and spatial arrangements on emotional responses could further refine design guidelines. Additionally, examining the influence of cultural and demographic factors on emotional responses to biodigital designs could provide a deeper understanding of user preferences across different populations